\documentclass{iau} 
\usepackage{graphicx}

\title[~~Dwarf galaxies and streams in M31] 
{The dynamics of Andromeda's dwarf galaxies and stellar streams}

\author[Michelle Collins et al.]   
{Michelle L. M. Collins$^1,$
R. Michael Rich$^2$, Rodrigo Ibata$^3$, Nicolas Martin$^3$ , Janet Preston$^1$\& The PAndAS collaboration}

\affiliation{$^1$Department of Physics, University of Surrey,
Guildford, GU2 7XH, UK \\ email: {\tt m.collins@surrey.ac.uk} \\[\affilskip]
$^2$Dept. of Astronomy UCLA, Los Angeles CA, USA \\
$^3$L'Observatoire de Strasbourg, Strasbourg, France \\}

\pubyear{2016}
\volume{321}  
\jname{Formation and evolution of galaxy outskirts}
\editors{A.C. Editor, B.D. Editor \& C.E. Editor, eds.}
\begin{document}

\maketitle

\begin{abstract}
As part of the Z-PAndAS Keck II DEIMOS survey of resolved stars in our neighboring galaxy, Andromeda (M31), we have built up a unique data set of measured velocities and chemistries for thousands of stars in the Andromeda stellar halo, particularly probing its rich and complex substructure. In this contribution, we will discuss the structural, dynamical and chemical properties of Andromeda's dwarf spheroidal galaxies, and how there is no observational evidence for a difference in the evolutionary histories of those found on and off M31's vast plane of satellites. We will also discuss a possible extension to the most significant merger event in M31 - the Giant Southern Stream - and how we can use this feature to refine our understanding of M31's mass profile, and its complex evolution.
\end{abstract}

\firstsection 
\section{Introduction}

The Pan-Andromeda Archaeological Survey (PAndAS) has mapped the stellar populations of our nearest neighbour, M31, from its central regions out to 150 kpc (\cite{mcconnachie09}). In addition to this impressive imaging  database, we have also obtained spectroscopic observations for tens of thousands of red giant branch stars (RGBs) throughout M31's disk, halo, satellites, and streams using Keck II DEIMOS. This follow-up survey - named Z-PAndAS - has been used to measure the masses of M31's vast satellite system (e.g. \cite{collins10,collins13}), uncover an extended and thick disk population in the system (\cite{ibata05,collins11}), map the M31 halo (\cite{chapman06})  and analyse a number of its streams and substructures (e.g., \cite{ibata04,chapman08}). In these proceedings, we focus on recent work comparing the properties of dwarf galaxies found on and off the vast plane of satellites in M31; as well as presenting kinematics for a possible extension to M31's most significant recent merger, the Giant Southern Stream.
\section{M31's plane of satellites - no evidence for differences in evolution on and off the plane}

To date, 31 dwarf spheroidal companions have been discovered around M31. Interestingly, $\sim50\%$ of these reside in a vast, thin plane. This plane is 400 kpc in diameter, has a scale height of only 14 kpc, and shows evidence for co-rotation (\cite{ibata13}). Such thin and extended planes are not common in nature (e.g., \cite{ibata14}), leading some to suggest that these dwarf galaxies have not formed within their own dark matter halos, but instead have formed tidally, out of dense streams of gas produced by a gas rich merger in M31 $\sim5-8$~ Gyrs ago (\cite{hammer14}). If this were the case, one might expect to observe significant differences in the properties of the on- and off-plane dwarf galaxies, as the on-plane dSphs would form much later than their off-plane counterparts, from pre-enriched gas, and  without a dark matter component.

\begin{figure}[b]
\begin{center}
 \includegraphics[width=1.7in]{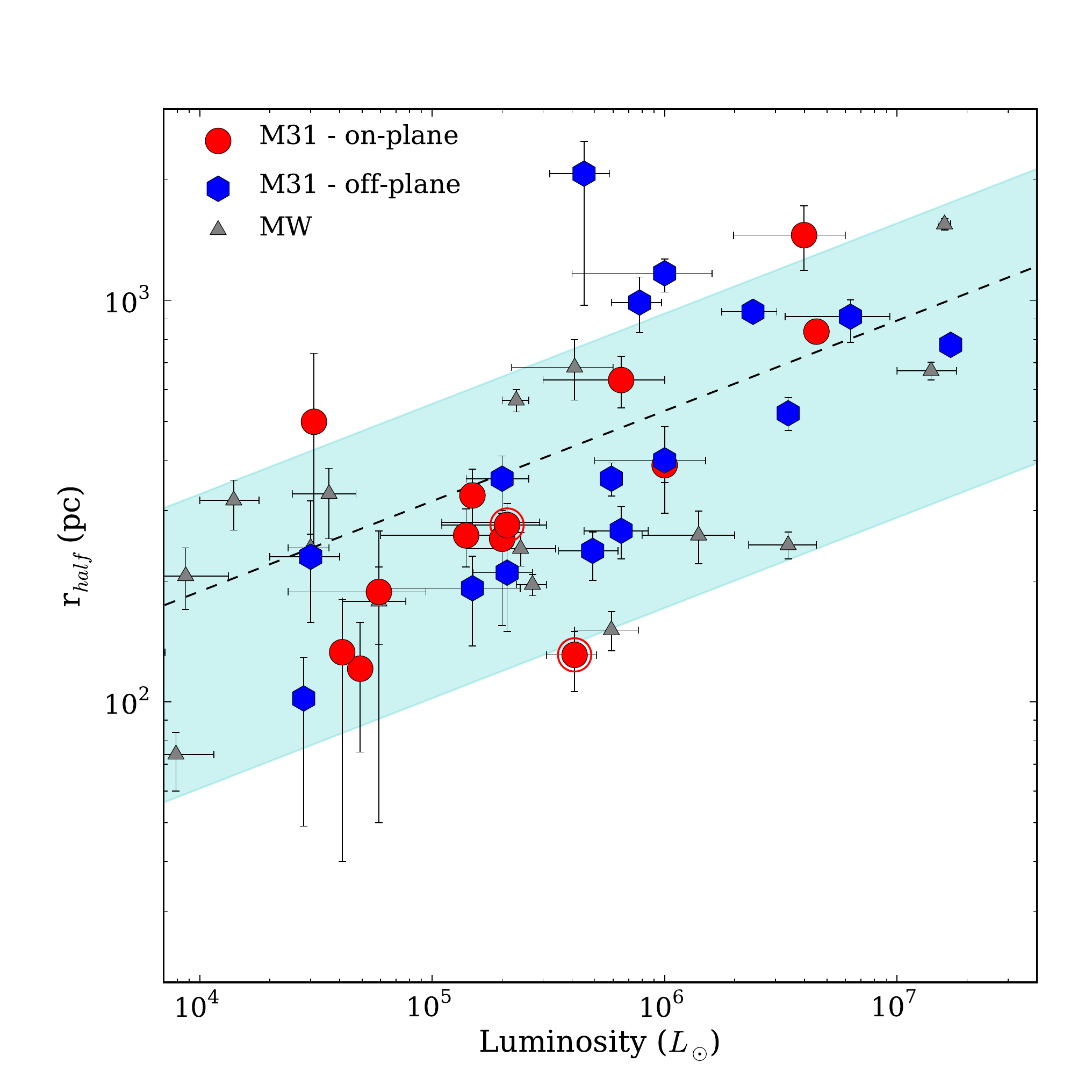} 
 \includegraphics[width=1.7in]{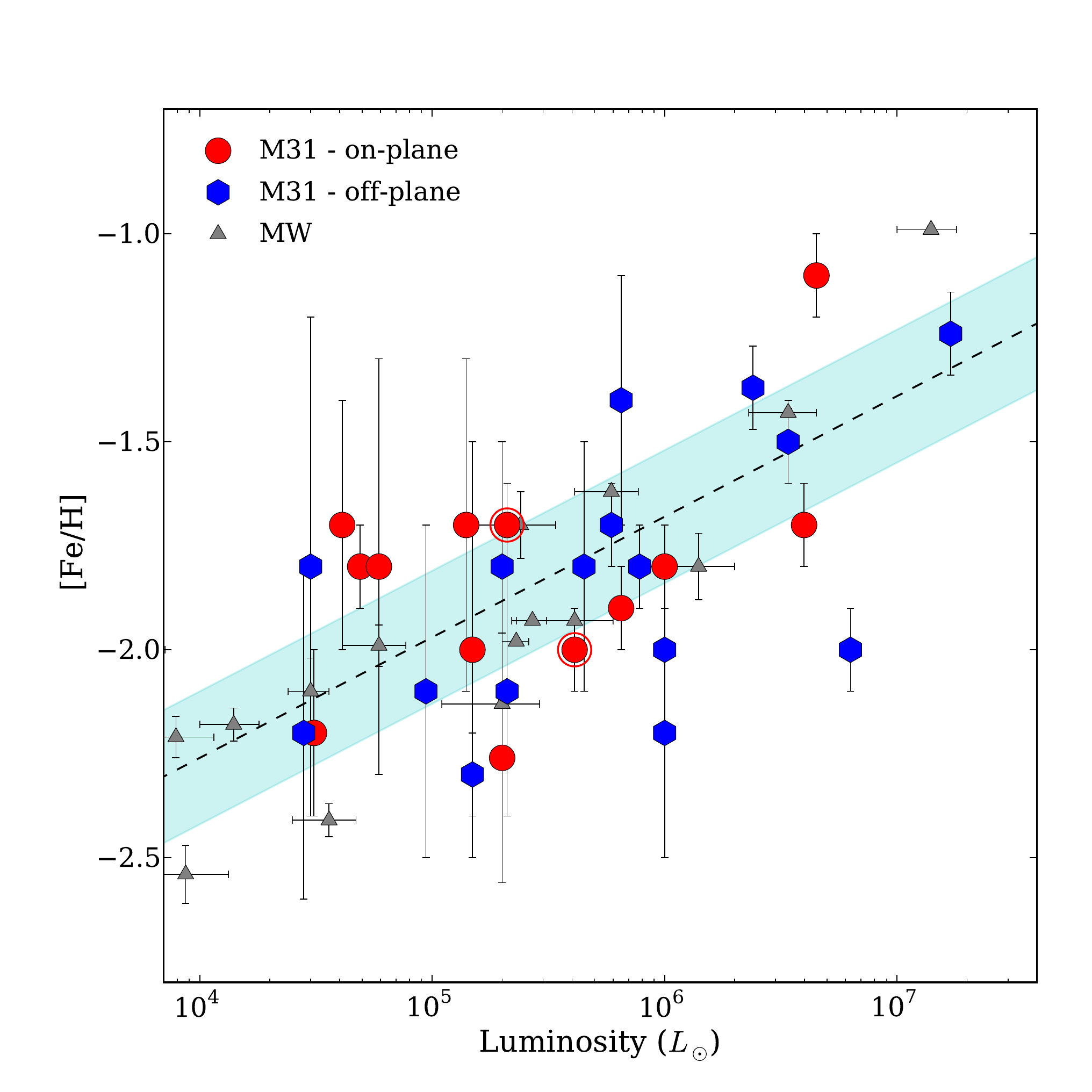} 
 \includegraphics[width=1.7in]{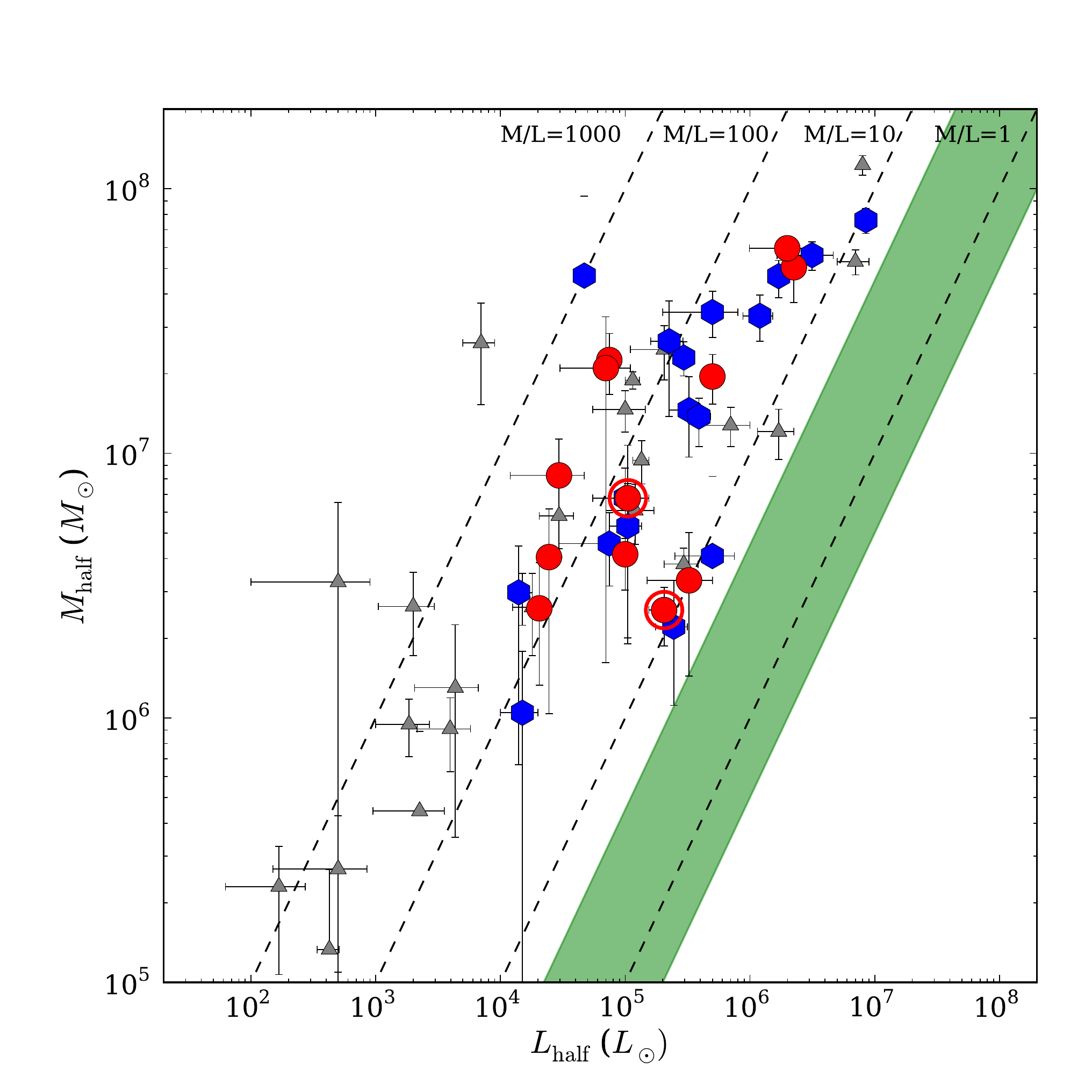} 
 \caption{Comparing the properties of on-plane (red circles) and off-plane (blue hexagons) M31 dSphs to established relationships for the Milky Way dSphs (grey triangles). {\bf Left:} Luminosity vs half-light radius for dSph galaxies. The size-luminosity relation from \cite{brasseur11} is shown as a dashed line, with the cyan band showing the $1\sigma$ scatter about this relationship. Both independent fits to the on- and off-plane dwarf galaxies, and KS tests show no statistical difference in how the on-plane, off-plane and Milky Way dSphs populate this parameter space. {\bf Centre:} The luminosity-metallicity relationship for Milky Way dSphs (dashed line and cyan band) as measured by \cite{kirby13}. Again, there is no statistical difference between how on-plane, off-plane and Milky Way dSphs populate this parameter space, implying they have all evolved under similar conditions. {\bf Right:} Mass within the half-light radius vs. luminosity for on-plane and off-plane M31 dSphs. Dashed lines represent lines of constant mass-to-light ratios of 1, 10, 100 and 1000 $M_\odot/L_\odot$ (from right to left). The green band is the region of parameter space inhabited by simple (baryonic) stellar systems. Both the on-plane and off-plane M31 dSphs have $M/L>10M_\odot/L_\odot$, implying they all formed within their own dark matter halos.}
   \label{fig1}
\end{center}
\end{figure}

In \cite{collins15}, we investigated the observable properties of all M31 dSphs using imaging and kinematic data taken for their individual stars. We compared the sizes, luminosities, chemistries and masses of the on- and off-plane dSphs with well established relations in these parameter spaces derived for Milky Way dSphs, including the size-luminosity relation (\cite{brasseur11}), the luminosity-metallicity relation (\cite{kirby13}), and the mass-luminosity relation (\cite{collins14}). In Fig.~\ref{fig1}, we show the results of these comparisons. In all cases, we found no statistical evidence for any differences between the on- and off-plane dSphs, nor from their Milky Way counterparts. As such there is {\it no evidence} that the dSphs within the M31 plane experienced a dramatically different evolutionary history from those that formed within their own dark matter halos off the plane. This finding is further bolstered by the measured star formation histories for M31 dSphs from \cite{weisz14}, where dSphs on and off the plane are found to host significant old stellar populations, and have similarly extended star formation histories.

 \vspace*{-0.5 cm}
\section{An extension to the Giant Southern Stream?}

\begin{figure}[h]
 \vspace*{-0.5 cm}
\begin{center}
 \includegraphics[width=1.9in, angle=0]{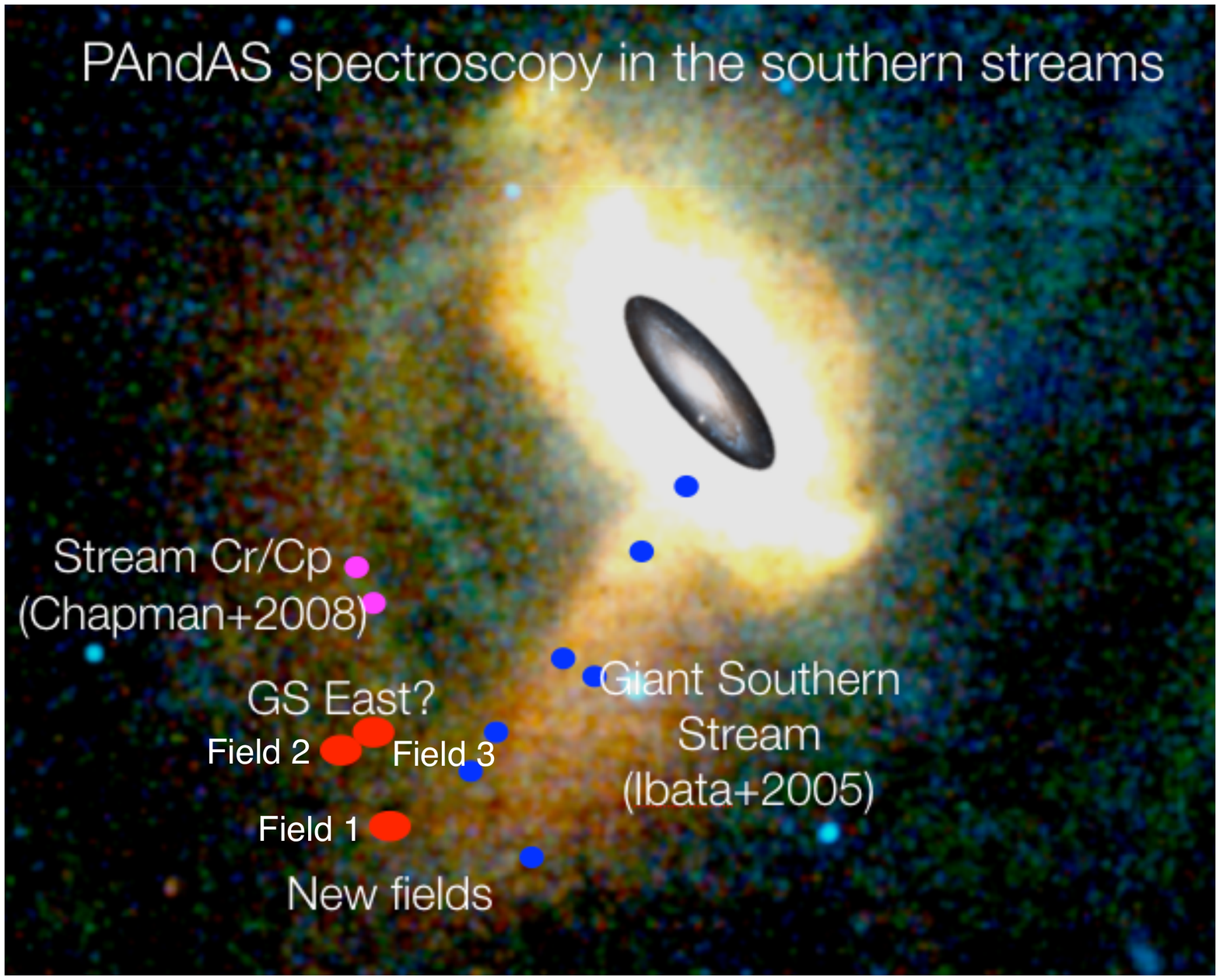} 
 \includegraphics[width=1.6in, angle=0]{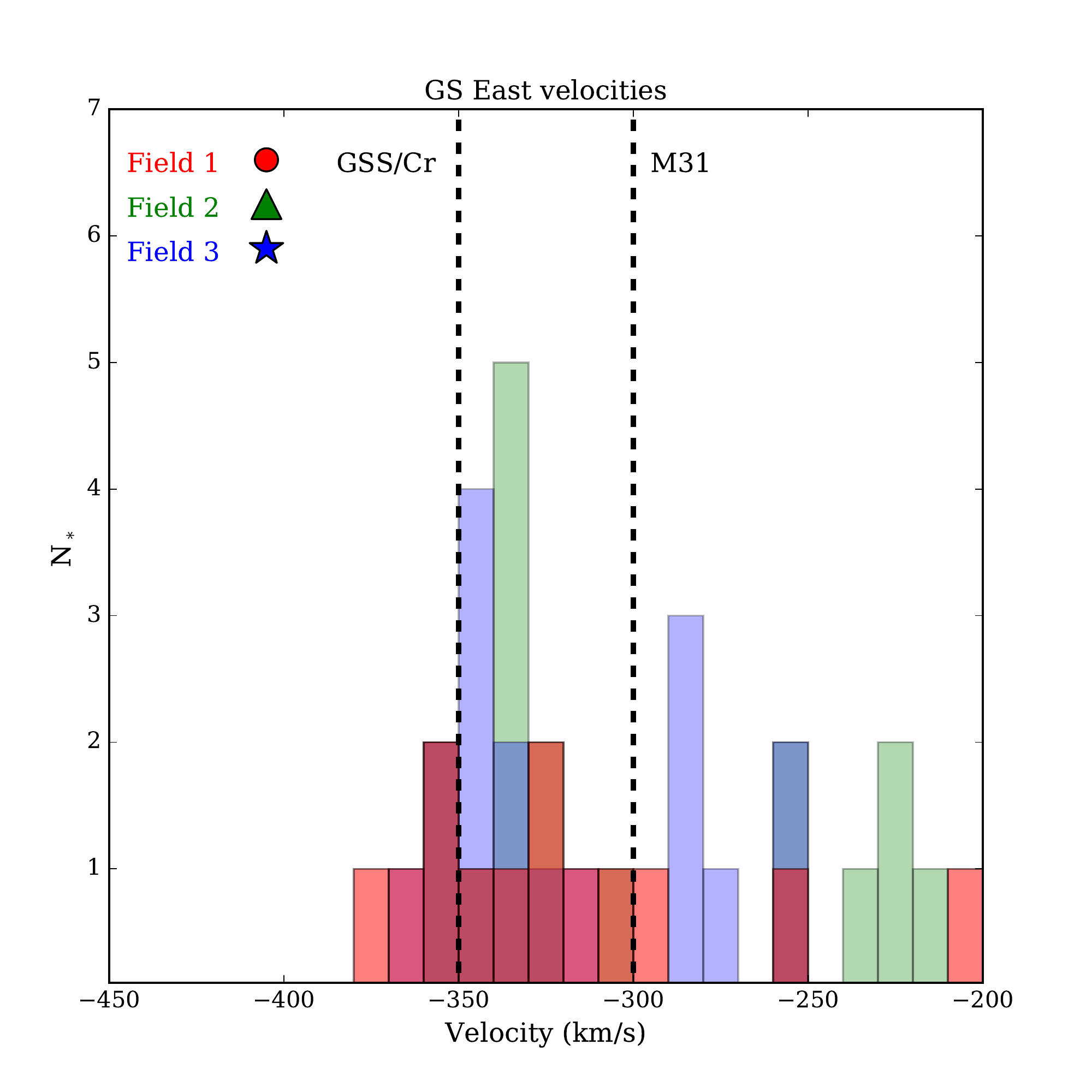} 
 \includegraphics[width=1.6in, angle=0]{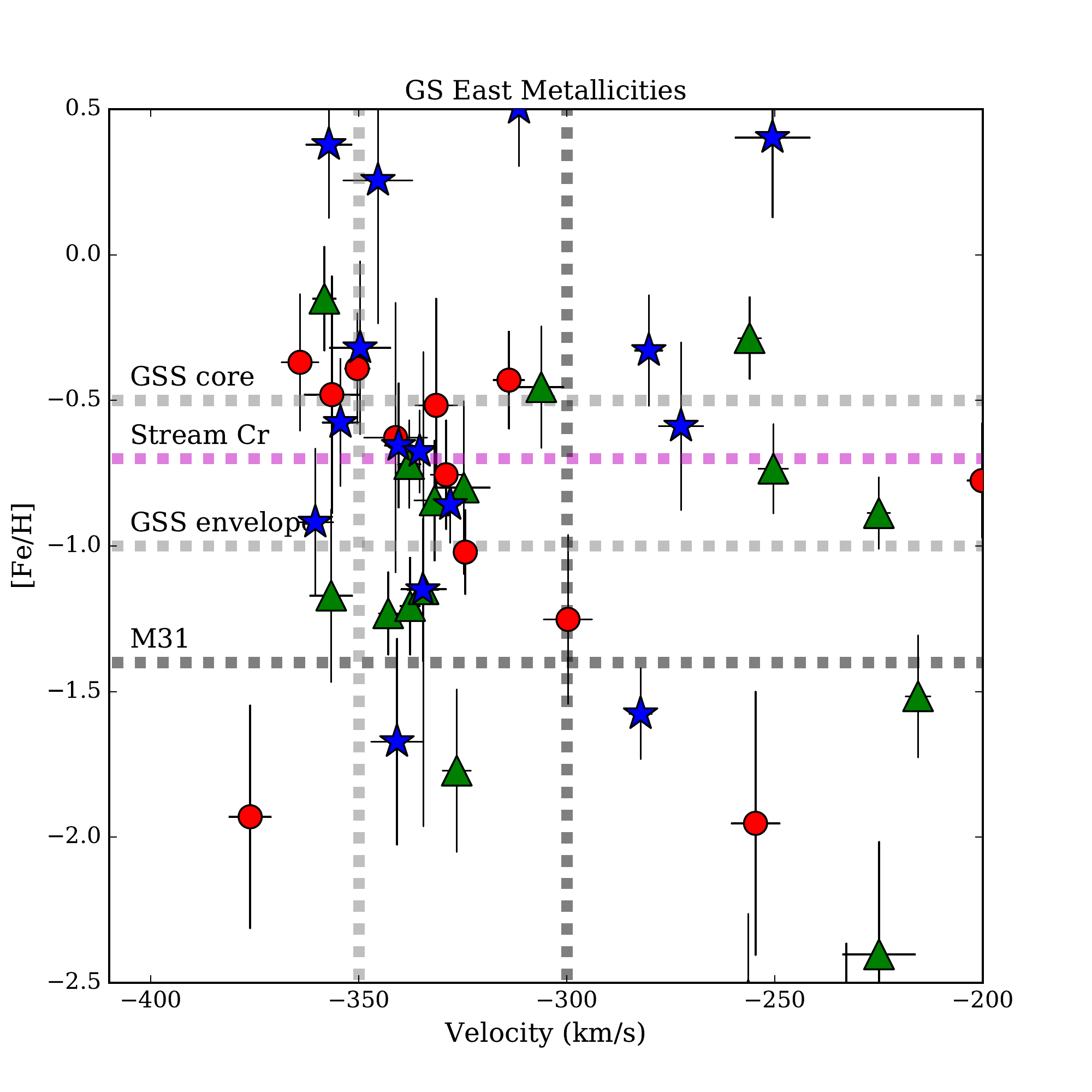} 
 \caption{{\bf Left:} PAndAS stellar density map of the GSS region of M31. Stars are colour-coded by their metallicity, with red equating to metal-enriched populations, and blue equating to metal-poor populations. The positions of our Keck DEIMOS spectroscopic fields in the GSS, Stream Cr and the new GSE extension are overlaid. {\bf Centre:} Kinematics of stars in our three GSE fields. Field 1 (red) is the closest to the GSS turn-off, with field 2 (green) and field 3 (blue) probing further along the GSE. In all three fields, there are a number of stars with velocities that are consistent with those of the outermost GSS field, and the stream Cr fields. {\bf Right:} Velocity vs. [Fe/H] for our 3 GSE fields. Many of the stars with GSS-like kinematics are also metal enriched, consistent with the enriched stars in both the GSS (grey dashed line) and stream Cr (dashed magenta line). }
   \label{fig2}
\end{center}
\end{figure}

The Giant Southern Stream is the most significant recent merger in the M31 system. It has been extensively studied both photometrically and spectroscopically  (e.g. \cite{mcconnachie04,ibata04,gilbert09}), and these data have been used to place constraints on the orbit of its unknown progenitor (e.g., \cite{fardal13}). Imaging from the PAndAS survey has recently shown that there may be an as yet un-studied portion of this stream, to the east of the GSS (GS East, see Fig.~\ref{fig2}). To ascertain whether this stream is truly part of the GSS, we have started a spectroscopic campaign with DEIMOS as part of the Z-PAndAS survey, so that we can establish its kinematics and chemistry. Our early results demonstrate that the GS East is composed of stars that have kinematics and metallicities consistent with both the GSS, and also with stream Cr, a metal-rich stream that runs tangential to the M31 disk (\cite{chapman08}). With more fields observed further along the GS East, we will be able to ascertain whether there exists a plausible orbit that can link this feature to the GSS (or Stream Cr), and if so, use this to better constrain the orbital history of the GSS, and the mass profile of M31. 

\end{document}